\documentclass[sigconf]{acmart}
\AtBeginDocument{%
  }

\usepackage{amsmath}

\usepackage{amssymb}
\usepackage{algorithmic}
\usepackage{graphicx}
\usepackage{textcomp}
\usepackage{xcolor}
\usepackage{booktabs}   
\usepackage{array}      
\usepackage{multirow}   
\usepackage{enumitem}

\newcommand{\red}[1]{\textcolor{black}{#1}}

\newcommand \tool{\texttt{\red{AFS-UI}}}

\setcopyright{none}
\renewcommand\footnotetextcopyrightpermission[1]{}

\begin{document}

\title{Everything is an Operational View: \\UI as a Projection over Context in Agentic Systems}

\author{Jieshan Chen}
\email{jieshan.chen@csiro.au}
\affiliation{%
  \institution{CSIRO}
  \country{Australia}
}

\author{Robert Mao}
\email{rob@arcblock.io}
\affiliation{%
  \institution{ArcBlock, Inc}
  \country{USA}
}

\author{Xiwei (Sherry) Xu}
\email{xiwei.xu@csiro.au}
\affiliation{%
  \institution{CSIRO}
  \country{Australia}
}

\author{Xuewu Gu}
\email{nate@arcblock.io}
\affiliation{%
  \institution{ArcBlock, Inc}
  \country{USA}
}

\author{Yechao Li}
\email{liyechao@arcblock.io}
\affiliation{%
  \institution{ArcBlock, Inc}
  \country{USA}
}

\author{Zhenchang Xing}
\email{zhenchang.xing@csiro.au}
\affiliation{%
  \institution{CSIRO}
  \country{Australia}
}

\renewcommand{\shortauthors}{Chen et al.}

\begin{abstract}
The emergence of generative AI has enabled Generative User Interfaces (GenUI) that dynamically construct interfaces from user intent, yet existing approaches rely on request–response pipelines that fragment context and limit transparency and control. This paper proposes AFS-UI, which reconceptualises UI as \emph{projections over context} built on an agentic file-system abstraction that organizes data, capabilities, and interaction state within a unified, persistent workspace. 
In particular, AFS-UI addresses key limitations in current systems, including heterogeneous resource discoverability, incompatible operation semantics, tangling projection and context, and rigid source composition and substitution, and limited evolvability. 
Within this framework, interfaces become dynamic, inspectable views over shared resources, supporting both human and agent interaction through a common operational surface. 
\end{abstract}



\maketitle

\section{Introduction}

Generative user interfaces (GenUI) are emerging as a new Human-AI interaction paradigm in which interface components and interaction flows are dynamically produced in response to user intent and evolving context~\cite{lee2025towards, lindley2026does}. Recent advances in large language models (LLMs) and agentic systems have enabled interfaces that synthesise UI elements on demand, support adaptive workflows, and integrate planning, execution, and interaction within a dynamic and malleable experience~\cite{cao2025generative, xu2025duetui, min2025malleable}. These developments challenge the traditional view of UI as a static rendering layer, positioning it instead as an integral part of agentic system behaviour.

Despite the progress of GenUI, existing approaches are using a request–response paradigm in which the UI is treated as a generated artefact from user intent~\cite{cao2025generative, a2ui, json-render, xu2025duetui, mcp-app, mcp-ui}. In such systems, context artefacts, including prompts, retrieved documents, memory, intermediate reasoning traces, and access policies, are managed fragmentarily within backend pipelines, while a separate UI layer consumes only final outputs. This separation leads to fragmented context, transient system state, and substantial glue code to bridge data, logic, and presentation. As a result, users have limited ability to inspect, trace, or influence the context that drives UI generation, and explanation and provenance are typically introduced as post-hoc augmentations rather than intrinsic properties of the system.

In the settings of agentic system that perform actions and coordinate over extended interaction, we envision that the \textit{UI should no longer be treated as a passive output surface, but an active participant in the workflow and context production and evolution}. Views are not only consumed by humans but also accessed and modified by agents~\cite{webmcp, schick2023toolformer, lu2025axis}. User interactions, such as navigation, selection, and configuration, directly influence system state and future behaviour. However, existing approaches remain tightly coupled to platform-specific UI technologies and provide limited abstractions for representing user intent, interaction history, and constraints as first-class contextual artefacts~\cite{vercelV0}.

In this work, we summarise common pipelines of existing generative UI systems and identify four key challenges through a motivating scenario. Building on prior work~\cite{xu2025contextagenticfileabstraction}, we extend a unified file-system abstraction to the UI layer by reconceptualising \textbf{UI as a projection over context}.
We introduce \red{Agentic File System for User Interface} (\tool{}), a \textbf{runtime substrate} that maintains a persistent, composable resource space, where heterogeneous resources are exposed as \textit{providers} under a uniform namespace with metadata, operations, provenance, and access control. The framework decouples system context from UI representation and rendering, treating interfaces as \textit{operational views} that bind to and evolve with system state.
This shifts GenUI from a \textit{transformation pipeline} to a \textit{persistent workspace}, where context, interaction history, and UI artefacts are maintained as evolvable entities. Users (humans and agents) can define, inspect, and modify this context, including UI specifications, through the projection.


Our contributions are as below: 
\begin{itemize}[leftmargin=14pt]
    \item We re-conceptualize generative UI as \textit{an integral projection over uniform context} in agentic systems, rather than as merely separate artifact generation within a request-response pipeline.
    \item We operationalise this conceptualisation by extending the agentic file system abstraction to the UI layer and propose the design of \tool{}. The resulting design organises context artefacts, UI specs, agent outputs, and interaction history within a unified, navigable namespace enriched with persistent metadata, provenance, and access control.
    
\end{itemize}

\begin{figure*}
    \centering
    \includegraphics[width=1\linewidth]{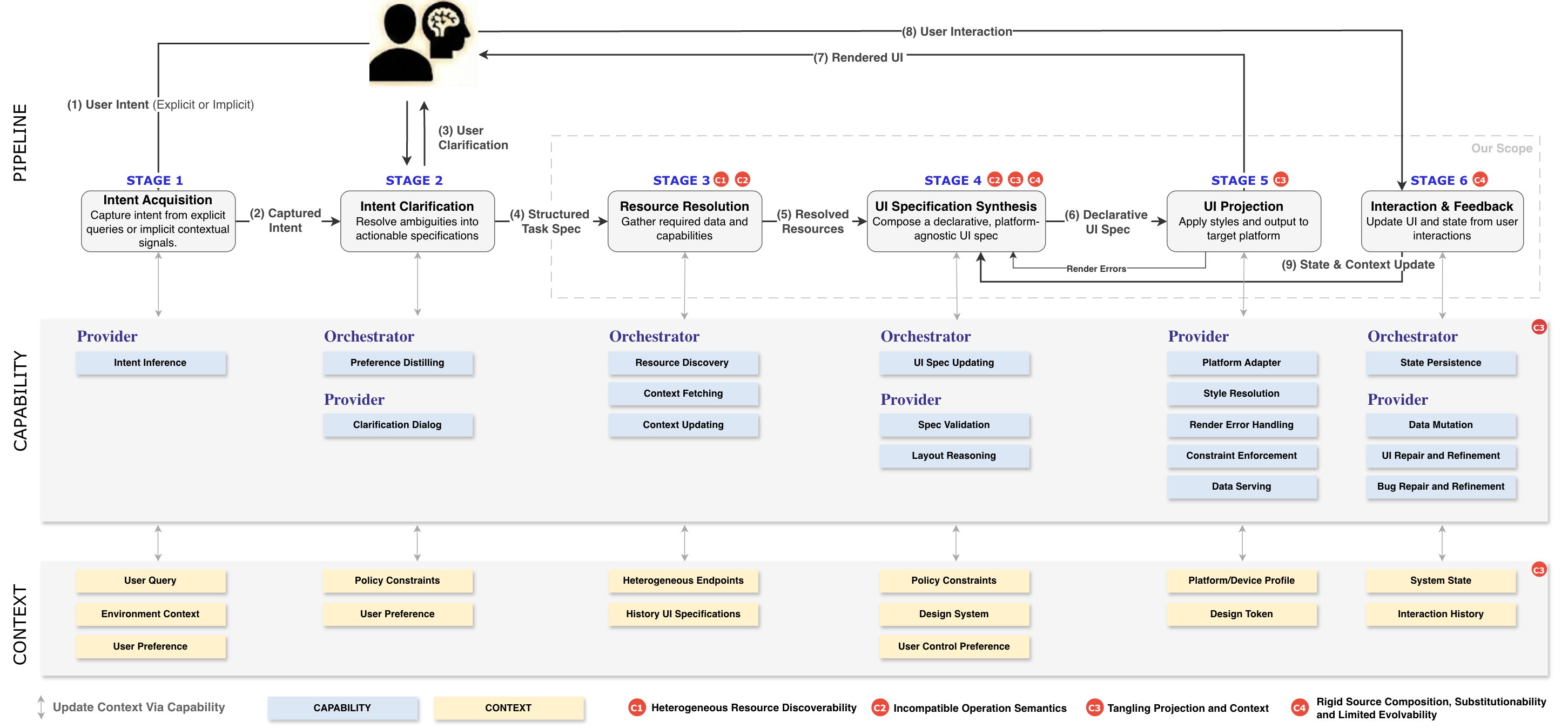}
    \caption{A generalized pipeline for generative UI systems. Six functional stages transform user intent into an interactive interface through an iterative cycle. Each stage depends on specific data and capabilities to fulfill its responsibilities. Provider and Orchestrator are concepts proposed in this work (Section~\ref{sec:roles}). The four challenges (C1–C4) identified in Section~2.2.1 are mapped to the stages where they arise (Section~\ref{sec:challenges}). 
    }
    \label{fig:genui_flow}
\end{figure*}

\section{Background and Motivation}
\label{sec:background}

In this section, we introduce a generalized pipeline for generative UI (GenUI) systems, identify its key stages, and use a motivating scenario to highlight existing challenges.

\subsection{A Generalized Pipeline of GenUI Systems}

Existing GenUI systems, despite differences in target domains and implementation strategies, can be understood through a common transformation pipeline. At a high level, these systems transform user intent into an interactive interface through a sequence of functional stages, each drawing on specific data and capabilities, while continuously incorporating feedback from users.

Figure~\ref{fig:genui_flow} illustrates this generalized workflow, which consists of six core stages: \textit{intent acquisition}, \textit{intent clarification}, \textit{resource resolution}, \textit{UI specification synthesis}, \textit{\red{UI projection}}, and \textit{interaction feedback}. Each stage depends on specific data (information that can be read, queried, or referenced) and capabilities (operations that can be invoked to perform actions) to fulfill its responsibilities, as shown at the bottom of Figure~\ref{fig:genui_flow}. Existing GenUI systems do not necessarily include all steps and may aggregate multiple steps into a single one.

\textbf{Stage 1: Intent acquisition} captures user goals from explicit inputs (e.g., natural language requests) or implicit signals derived from environment context, prior interactions, or user preferences. These signals provide the initial entry point for UI generation.

\textbf{Stage 2: Intent clarification} refines and structures the captured intent into a precise, actionable task specification. This stage may involve resolving ambiguities, augmenting missing information through clarification dialogue, and consulting user preferences and interaction history to interpret intent in context.

\textbf{Stage 3: Resource resolution} identifies and prepares the data and capabilities required to support the task. This involves discovering which resources (e.g., databases, user preferences, tools, services, design tokens) are available and fetching what is needed for downstream UI design and construction.

\textbf{Stage 4: UI specification synthesis} determines how the resolved resources should be organized and presented. Rather than directly generating implementation code, the system constructs a declarative, platform-agnostic UI specification that captures component structure, interaction logic, and data bindings. This stage draws on policy constraints, design tokens, and user control preferences, and may apply specification validation to ensure the specification is well-formed before rendering. In some GenUI systems, the final rendering code may be generated directly, or any intermediate specification, if present, remains implicit and opaque to the user~\cite{vercelV0, bolt, mcp-app, leviathan2025generative}.

\textbf{Stage 5: \red{UI Projection}} instantiates the specification into a platform-specific interface by applying design system styles, resolving platform and device constraints, and performing render-time error handling. This stage focuses on visual execution and platform adaptation, without introducing additional semantic decisions. During this stage, rendering errors may occur, which are propagated back to the previous stage for correction.

\textbf{Stage 6: Interaction feedback} closes the loop by incorporating both interface refinement and state updates. Users may explicitly request changes to the generated UI, triggering re-generation through UI repair and refinement, or interact with it to perform tasks, producing updates to underlying data and system state. These updates are persisted and inform subsequent UI generation cycles.

Together, these stages form an iterative pipeline in which UI generation is not a one-off process, but a continuous cycle between user intent, system resources, and interface representations. As shown in Figure~\ref{fig:genui_flow}, each stage relies on overlapping yet distinct sets of data and capabilities, a structural observation that motivates our analysis of existing systems' limitations in the following sections.

In general, Stages 1 and 2 focus on requirements engineering, Stages 3–5 form the technical core of a generative UI system, and Stage 6 addresses evolvability and human–AI co-design. \textit{This paper focuses primarily on Stages 3–5 and partially on Stage 6}, introducing a system-level substrate for GenUI as a unified operational surface that integrates UI generation with the management of underlying data, providers, and context.
The human–AI interaction aspects of Stages 1, 2, and 6 are left as future work.

\subsection{Motivating Scenario}
\label{sec:motivation}

Consider a project manager at a mid-sized software company who uses multiple communication platforms daily: \emph{email} for formal correspondence with clients, \emph{Slack} for team discussions and project channels, \emph{Microsoft Teams} for organization-wide announcements and meetings, and \emph{WhatsApp} for quick coordination with remote contractors. 
Her messages about the same project are scattered across all four apps, forcing her to context-switch constantly just to piece together a complete picture. Frustrated, she turns to an AI assistant and asks:
\emph{``Show me all my recent messages in one place, newest first. Tag each one with where it came from.''}
Fulfilling this request is far from straightforward. Her organization enforces strict security and compliance policies. Some client communications contain sensitive information that cannot be freely aggregated, stored, or processed by external AI services. As a result, any unified view must respect access control boundaries, selectively retrieve content, and potentially apply summarization while preserving provenance. In addition, the company follows a unified design system that constrains how information can be presented. Furthermore, she personally prefers concise summaries for Slack and WhatsApp messages but more detailed views for emails and formal documents.
 
From her perspective, the request is simple: a single, chronological feed with clear source attribution. Yet behind the scenes, the system must  
(1) retrieve messages from four independent platforms, each with its own API, authentication scheme, and data model;
(2) reconcile fundamentally different notions of ``a message'' across these platforms (an email thread, a Slack channel post, a Teams announcement, an encrypted chat message in WhatsApp); 
(3) perform entity matching and resolution to associate messages that relate to the same project across sources despite inconsistent identifiers, naming conventions, or implicit references;
and (4) normalize timestamps and metadata sufficiently to interleave messages from all sources into a single coherent timeline. These requirements demand capabilities beyond generating a nice-looking user interface, requiring cross-source integration, semantic alignment, and metadata-aware reasoning.
 
\subsubsection{Challenges}
\label{sec:challenges}
This scenario illustrates typical generative UI tasks in which a GenUI agent must assemble an interactive interface over multiple heterogeneous data sources. We identify \red{four} challenges that arise in 
the tasks within Stages 3-6 in Figure~\ref{fig:genui_flow}.

\textbf{C1: Heterogeneous Resource Discoverability (\textit{Stage 3}).} 
Assembling the interface requires the agent to locate and access data scattered across independent platforms. Without a shared workspace in which all relevant sources are uniformly discoverable, the agent cannot systematically determine what resources are available for UI assembly, while hardcoding resource bindings inevitably degrades evolvability. 
Current GenUI systems sidestep this challenge by assuming that all relevant resources are already available in a ready-to-use format with well-defined data schemas, rather than discovered and assembled at runtime~\cite{cao2025generative, xu2025duetui}.

\textbf{C2: Incompatible Operation Semantics (\textit{Stage 3}).}
Even when all sources are reachable (e.g., exposed via MCP endpoints), each exposes a heterogeneous API with distinct function names, parameter structures, and invocation patterns. Even when well-documented, these differences consume context and increase cognitive load, forcing the agent to reason over multiple API contracts within a single generation pass and increasing the likelihood of incorrect or incomplete invocations.
While manageable for a small number of sources, this burden grows with each additional platform, increasing the likelihood of malformed invocations or incomplete data retrieval, as the model's context is consumed by API-specific boilerplate (most irrelevant to the tasks at hand) rather than task-level reasoning~\cite{anthropicCodeExecutionBlog}.

\textbf{C3: Tangling Projection and Context (\textit{Stage 4 \& 5}).}
Interface generation involves three distinct concerns: data sources (backend context), which supply the underlying information; the interface specification (frontend context), which defines what should be presented (e.g., components and content structure); and the projection layer, which determines how the underlying information and interface specification is rendered (e.g., layout, styling, and interaction patterns). 

Even though these concerns are logically separable, existing systems often entangle them within a single generation process. Data is frequently embedded, co-generated, or explicitly orchestrated together with the interface, forcing the agent to reason over content, structure, and presentation simultaneously within a shared context~\cite{a2ui, json-render, cao2025generative, xu2025duetui}.
This entanglement introduces implicit dependencies between layers, where changes in data, structure, or rendering require coordinated modifications across the entire interface. As complexity grows, this coupling reduces modularity and makes interface generation less systematic, less predictable, and more error-prone.

\textbf{C4: Rigid Source Composition and Substitution, and Limited Evolvability (\textit{Stage 4 \& 6}).}  
Real-world systems operate over evolving and heterogeneous data sources. Services change, APIs are replaced, and users frequently switch between equivalent platforms (e.g., Slack to Discord). These changes do not alter the underlying user task, yet they often invalidate existing integrations. 

Current systems integrate data sources through platform-specific schemas, APIs, and interaction patterns, resulting in compositions that are tightly coupled to particular sources. Substituting one source for another requires modifying tool definitions, response parsers, or even the generation pipeline. 
This rigidity extends to the interface level. Generated interfaces are tightly bound to specific data sources and usage contexts, making them difficult to reuse, recombine, or extend across tasks. For example, augmenting an existing interface with additional functionality, such as incorporating a data dashboard alongside a message view, typically requires restructuring or regenerating the interface, rather than incrementally composing new components.

As a result, composition, substitution, and evolvability are not treated as first-class capabilities. Changes in data sources or interface requirements propagate as implementation-level modifications rather than configuration or binding adjustments. Consequently, supporting multi-source aggregation, cross-provider interaction, runtime source switching, or stackable user interfaces all requires substantial engineering effort outside the system’s core abstractions.

\section{System Architecture}
\label{sec:architecture}

\begin{figure*}
    \centering
    \includegraphics[width=0.8\textwidth]{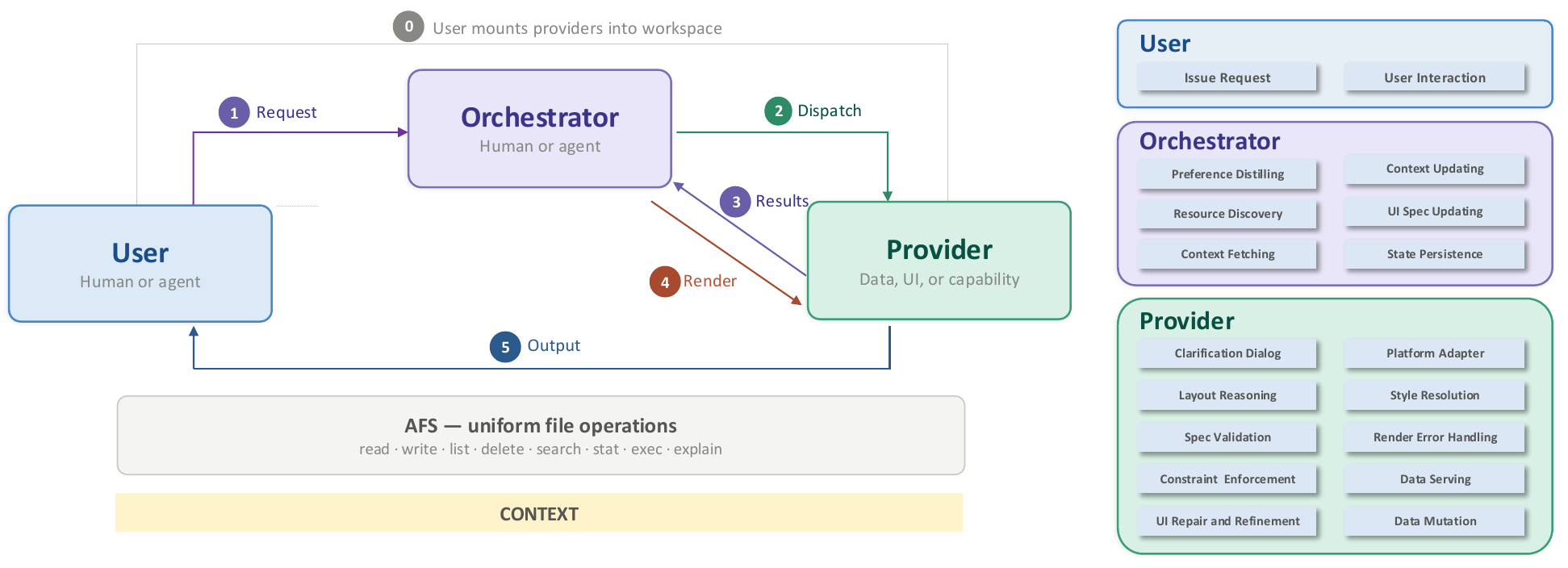}
    \vspace{-10pt}
    \caption{The interaction flow between three roles, namely users, orchestrator, and providers, in our system. Capabilities for each role are shown on the right, corresponding to Figure~\ref{fig:genui_flow}.
    }
    \label{fig:afs_roles}
\end{figure*}

\subsection{Reconceputalizing GenUI: From Transformation to Semantic Task Space}

Recent generative UI systems enable large language models to dynamically produce user interfaces and/or code from natural language instructions and/or design images~\cite{vercelV0, bolt, lovable}. 
However, they mainly treat the UI as a \textbf{model-generated artifact}, re-generated or updated in response to evolving inputs.
UI state is typically scoped to individual inputs or input updates,
while coordination across multiple agents, data sources, and persistent workflows requires additional orchestration logic outside the UI generation framework. This limitation becomes particularly apparent in emerging \textbf{agentic computing environments}, where multiple agents and human continuously interact with shared data, tools, and long-lived task contexts. In such settings, the UI is not merely an output of a single model generation, but as \textbf{a projection of evolving user input, system state and interaction context}.

\subsection{Design Principles}
\label{sec:principles}
To address the challenges in Section~\ref{sec:challenges}, we formulate a set of design principles (P) that define the invariants of our modeling of GenUI systems. These principles guide how resources are represented, accessed, and composed, and serve as the foundation for the architecture described in the following sections.

\textbf{P0. Shared Semantic Workplace.}
All agents, data sources, and interfaces operate within a unified, persistent workspace that serves as the common ground for resource discovery, state management, and interface projection. Rather than coordinating across isolated systems through external orchestration, the workspace provides a single environment in which resources coexist, relationships are explicit, and system state is continuously maintained. This shared foundation underpins all subsequent principles and is the basis for addressing the challenges identified above.

\textbf{P1. Uniform Addressability.} 
Within the shared workspace, all resources, including data sources, tool endpoints, UI specification, and generated artifacts, are represented under a uniform addressing scheme in which every resource is a path. 
Rather than accessing resources through heterogeneous APIs, the system treats them as uniformly addressable entities that can be uniquely identified, referenced, and retrieved via a consistent addressing mechanism.
This unifies access under a single abstraction, allowing the system to operate over a coherent resource space rather than a collection of disparate endpoints.
It also enables UI generation to be framed as manipulation of persistent system state, rather than a sequence of transient API calls.
Resource discovery is thus transformed from a per-source integration problem into a structured traversal problem, directly addressing the discoverability challenge (C1).

\textbf{P2. Uniform Operation Semantics.}
All system capabilities are exposed through a standardized set of operation semantics, regardless of their underlying implementations.
P2 addresses C2 (Incompatible Operation Semantics) by constraining all interactions to a small set of standardized operation types. Rather than reasoning about arbitrary function signatures, the agent operates over uniform interaction patterns, regardless of the underlying implementation of each capability. This shifts the focus from API-specific usage to capability-level semantics.
This design decouples capability definition from orchestration, allowing capabilities to be reused, extended, or substituted without requiring the agent to relearn new invocation patterns. By reducing variability in interaction semantics, the system improves navigability and robustness.

\textbf{P3. Separation of Projection and Context.}
The three concerns identified in C3, data access, interface specification, and visual projection, are explicitly separated into independent, modifiable layers. Changes in one layer do not require coordinated modifications in the others, directly addressing C3.

\textbf{P4. System-Wide Composability.}
All resources and capabilities are integrated through a plug-in mechanism into a shared namespace, enabling flexible composition across providers.
Each provider can be independently attached, detached, or replaced to the shared semantic workspace at runtime, without affecting the overall system structure. This allows multiple data sources, capabilities, and UI overlays to be combined dynamically, forming a unified interface over heterogeneous resources.
By treating providers as composable units within a common namespace, the system supports source substitution and multi-source aggregation without requiring changes to the generation logic, directly addressing the composition, substitution and evolvability challenge (C4).

\textbf{P5. UI as Projected Context.}
User interfaces are projections over the shared workspace, not standalone artifacts. A UI does not encode its own isolated state; it is derived from the current state of workspace resources through a projection mechanism that determines what is presented and how. 
By treating UI as a projection, the same underlying resources can be represented across different modalities, devices, or interaction contexts without redefining the interface logic. 
This design aligns interface representation with system state and avoids implicit duplication of information.
It also enables UI to be dynamically recomputed as the resource  evolves, addressing the evolution challenge (C4).

\subsection{Design Overview}
This section presents the system architecture that operationalizes the principles from Section~\ref{sec:principles}. The architecture comprises two complementary layers: a \emph{context layer} that unifies all system resources under a single addressable, composable namespace with the same operational primitives, and a \emph{projection layer} that derives interactive user interfaces from that namespace. Three roles, user, orchestrator and provider, interact within this architecture through the context layer's uniform semantics.

The context layer is built on the \emph{Agentic File System} (AFS)~\cite{xu2025contextagenticfileabstraction}, a virtual file-system abstraction that realizes the shared semantic workspace (P0).
All resources, data sources, agent capabilities, UI specifications, and generated artifacts, are encapsulated as \emph{providers}: self-contained units that are explicitly mounted into a hierarchical namespace. Within this namespace, every resource is addressable via a uniform path-based scheme (P1), and every provider exposes a standardized set of operations such as read, write, and exec (P2). Data bindings reference paths rather than concrete instances, and UI primitives are separated from rendering logic (P3), so providers can be freely attached, detached, or replaced at runtime without disrupting existing configurations (P4).

The projection layer is realized by \tool{}, a dedicated provider mounted into the same namespace (e.g., \texttt{/ui}). AFS-UI treats the user interface as a projection over workspace state (P5) rather than an independent artifact: the orchestrator traverses the namespace to discover relevant resources, retrieves data through standardized operations, and writes a UI specification to a session of AFS-UI, which renders it into an interactive interface (e.g., TTY or Web).

\subsection{Roles and Interaction Model}
\label{sec:roles}
The interaction model defines three functional roles --- user, orchestrator, and provider --- that are not tied to fixed components. A single instance (e.g., an LLM) may assume different roles at different stages of task execution.

The \textbf{user} initiates tasks and consumes results. This role may be filled by a human or an AI agent. It is responsible for configuring the workspace by mounting providers, both data sources and UI capabilities, into the AFS namespace, and for driving the interaction loop by issuing requests and acting on outputs. The \textbf{provider} supplies capabilities. Each provider encapsulates a resource, a certain capability, data source, or rendering engine, behind the uniform AFS interface. Providers operate independently and are agnostic to their callers. The \textbf{orchestrator} coordinates task execution. Upon receiving a request, it discovers mounted providers, decomposes the request into sub-tasks, dispatches them through standardized operations, and assembles results. This role is typically fulfilled by an LLM, but may also be assumed by the user or another agent.

As shown in Figure~\ref{fig:afs_roles}, interaction proceeds as a recurring cycle. The user first configures the workspace by mounting the required providers (step 0), then issues a request to the orchestrator (step 1). The orchestrator discovers relevant providers, decomposes the request, and dispatches sub-tasks (step 2). Providers execute and return results (step 3). The orchestrator composes the outputs and writes them to the UI provider for rendering (step 4), which delivers the interface to the user (step 5). User interaction with the rendered output (e.g., clicking, filtering, replying) triggers the next cycle.


\subsection{Context and Projection Layers}
This section details the two layers introduced in the design overview: the context layer (Section~\ref{sec:afs}), which summarizes the AFS abstraction inherited from \cite{xu2025contextagenticfileabstraction}, and the projection layer (Section~\ref{sec:afs-ui}), which constitutes the technical contribution of this paper. We use the motivating scenario from Section~\ref{sec:motivation} to illustrate key mechanisms throughout.

\subsubsection{Context Layer: Agentic File System}
\label{sec:afs}
The context layer is the Agentic File System (AFS)~\cite{xu2025contextagenticfileabstraction}, a virtual file-system abstraction in which every resource is a path-addressable \emph{entry} in a shared namespace. AFS ships with data-oriented providers (e.g., filesystem, JSON, SQLite).
 
\textbf{Entry.}
An entry is the atomic unit of the namespace. Each entry carries three facets: \emph{content} (the payload, a message body, a structured record, or a binary attachment), \emph{metadata} (the kind, content type, timestamp, provenance, and access-control annotations of the entry), and \emph{actions} (executable operations available on the entry, each with a name, input schema, and severity level). This tripartite structure means that an agent encountering any path, whether it resolves to an email thread, a Slack channel post, or a WhatsApp chat, can inspect what the resource is, what it contains, and what can be done with it, through a single uniform representation (P0).

\textbf{Operations.}
All entries are manipulated through a fixed set of eight operations: \texttt{read}, \texttt{write}, \texttt{list}, \texttt{delete}, \texttt{search}, \texttt{stat}, \texttt{exec}, and \texttt{explain}. The first six mirror conventional filesystem semantics. The latter two extend the abstraction to agentic settings: \texttt{exec} invokes an action declared on an entry (e.g., triggering an LLM summarisation or sending a reply), and \texttt{explain} returns a human- and agent-readable self-description of what exists at a given path and what operations are available. Together, \texttt{exec} and \texttt{explain} close the gap between passive data access and active capability negotiation and invocation that traditional filesystems leave open (P2), while enabling agents to discover and reason about unfamiliar resources without external documentation (P1).
 
\textbf{Mount-based composition.}
Heterogeneous data sources are integrated into the namespace through a mount mechanism. Each \emph{provider}, an independently developed module that implements a subset of the eight operations, is mounted at a designated path. At runtime, the system resolves an operation on any path by dispatching to the provider whose mount point is the longest prefix match. In our scenario, four messaging backends and an AI capability are composed into a single workspace:
\texttt{/mail} $\to$ \textsf{EmailProvider}, 
\texttt{/slack} $\to$ \textsf{SlackProvider}, 
\texttt{/teams} $\to$ \textsf{TeamsProvider}, 
\texttt{/wa} $\to$ \textsf{WhatsAppProvider}, 
\texttt{/ai} $\to$ \textsf{LLMProvider}, and, introduced in this paper (Section~\ref{sec:afs-ui}), 
\texttt{/ui} $\to$ \textsf{UIProvider}.
The agent calls \texttt{list("/mail/inbox")} and \texttt{list("/slack/channels/project-x")} using the same operation and the same entry schema, even though the underlying APIs, authentication schemes, and data models are entirely different. Each provider translates its platform-specific artefacts (email threads, channel posts, Teams announcements, encrypted WhatsApp messages) into the common entry representation, absorbing the reconciliation burden that would otherwise fall on the agent. Providers can be mounted, unmounted, or replaced at runtime without affecting the rest of the namespace (P4). 

\textbf{Policies.}
Security constraints are expressed within the same framework. Each provider declares an \emph{access mode} (e.g.\ \texttt{readonly}) and entries carry metadata including sensitivity annotations. The organisation's compliance policy-certain client emails must not be sent to external AI services-is enforced at the mount level: the email provider can restrict which entries are visible to downstream \texttt{exec} calls, without requiring the agent to implement policy logic.
 
\subsubsection{Projection Layer: UI Provider}
\label{sec:afs-ui}
 
The projection layer is realised by a dedicated UI provider (\textsf{AFS-UI}), mounted at a path (e.g., \texttt{/ui}) that lets agents construct and update user interfaces through ordinary AFS operations.
Writing a UI specification to a session path causes the connected device (i.e., the client), a browser, a terminal, or a mobile app, to render it; user events flow back as \texttt{exec} calls into the namespace. There is no separate UI SDK, the agent interacts with the interface through the same \texttt{read}, \texttt{write}, and \texttt{exec} operations it uses for any other resources. 

The data format \textsf{AFS-UI} speaks is the Agent UI Protocol (AUP), a declarative specification for describing UIs as semantic node trees. AUP is a standalone schema package with no runtime dependencies but pure types and validation, ensuring that the interface description is transport- and renderer-independent.

\textbf{Semantic node tree.}
An agent constructs a UI by building a tree of \emph{semantic nodes}. Each node carries a \emph{type} drawn from a vocabulary of built-in primitives (e.g.\ \texttt{text}, \texttt{input}, \texttt{action}, \texttt{table}, \texttt{chart}, \texttt{media}, \texttt{nav}), type-specific \emph{properties}, optional \emph{event bindings}, and an ordered list of \emph{children}. Node types describe \emph{what} to present, not \emph{how} to render it. For the unified inbox, the agent emits \texttt{view} nodes, each containing a \texttt{text} node for the message body and metadata props for source, timestamp, and sender, without specifying fonts, colours, or layout.

\textbf{Data binding.}
AUP nodes are connected to the AFS namespace through two binding mechanisms. A read-only binding (\texttt{src}) associates a node with an AFS path from which its display content is derived: a message card with \texttt{src:\,"/slack/channels/project-x/msg-42"} projects that Slack post without duplicating its data into the UI specification. A read-write binding (\texttt{bind}) additionally writes user-modified values back to the bound path. 
The interface maintains no copy of the data; it references the authoritative resource in the namespace (P5). When a new message arrives, the projection is recomputed from the source path; when the user edits a draft reply in a bound input node, the change propagates back to the source of truth. Because no copy exists, the projection cannot diverge from its source.

Different from frameworks such as React and A2UI~\cite{a2ui}, data is bound to an AFS path rather than to in-memory instances. 
Updates flow through the underlying provider, eliminating the event triggers and glue code that synchronize state copies within an application.

\textbf{Event dispatch.}
Each AUP node can declare event bindings that map user interactions to AFS \texttt{exec} calls. For example, a message card may carry:
\smallskip
{\small
\begin{verbatim}
"events": {
  "click": {
    "exec": "/ai/summarize",
    "args": { "path": "/mail/inbox/msg-7" }
  }
}
\end{verbatim}
}
\smallskip
\noindent When the user clicks this card, the client (i.e., the browser) forwards the event to the provider through the session's WebSocket connection. The UI provider resolves the node's binding and dispatches \texttt{exec("/ai/summarize", \{path: "/mail/inbox/msg-7"\})} into the namespace, the same operation an agent would invoke programmatically. There is no separate event bus or callback registry; human interactions and agent-initiated actions flow through the same operational interface (P2). 
The exec result can in turn be written to a namespace path, which a bound node then projects as an updated display.

\textbf{Session lifecycle and incremental patching.}
The orchestrator writes a completed AUP tree to the UI provider through an \texttt{exec} call. The provider opens a \emph{session}, a stateful connection to a client (e.g., a browser), pushes the tree over WebSocket; the browser renders it and receives user events. 
The client acts purely as a rendering surface while all logic stays within the namespace.
Subsequent changes are delivered as node-level patches: when a resource changes, only the nodes bound to it are re-rendered, while the rest of the tree remains untouched. As a result, transient state that exists only on the client — a reply the user has half-typed, the scroll position in a long feed — persists across updates.

\textbf{Surface composability.}
A \texttt{surface} is a built-in AUP primitive that embeds an independent rendering context within a node tree. Its \texttt{src} accepts an AFS path or a WebSocket URL. 
Given a WebSocket URL, the client establishes a second AUP session to the remote endpoint, rendered within the bounds of the enclosing node. Each embedded surface maintains its own session, event routing, and tree versioning.
Given an AFS path, the surface introspects the target (via \texttt{stat}, \texttt{list}, and \texttt{read}) and selects a rendering strategy: directories are rendered as navigable listings, files as detail views, and paths carrying provider-authored UI recipes (under a conventional \texttt{.aup/} directory) are rendered according to those recipes. 
The enclosing surface constrains the embedded content through capability negotiation, restricting which primitive types the inner tree may use. Primitives outside the whitelist are replaced with fallback views before rendering. 
This extends security constraints to the UI layer.

Surface thus lifts mount-based composability to the projection layer (P4). Just as the namespace composes heterogeneous data providers, \texttt{surface} composes heterogeneous UI sources, local agent-generated trees, remote agent sessions, and auto-projected AFS resources, into a single interface. A dashboard, for example, can embed a surface bound to a remote monitoring agent alongside a surface bound to \texttt{/db/metrics}, each operating as an independent projection with its own update lifecycle, without the orchestrating agent managing their internal state.

\textbf{Device-adaptive projection.}
At connection time, the client declares a capability profile describing its platform, form factor, and per-primitive rendering support (\texttt{native}, \texttt{partial}, or \texttt{unsupported}). Before transmitting the tree, the server applies a degradation pass: unsupported primitives are replaced along predefined fallback chains (e.g.\ \texttt{chart} $\to$ \texttt{table} $\to$ \texttt{text}). Because the AUP tree describes semantic intent rather than visual layout, the same inbox specification renders as a card-based feed on a desktop browser, a compact list on mobile, or a plain-text summary on a terminal, without modification to the interaction logic (P3).

\subsubsection{End-to-End Data Flow}
\label{sec:e2e}
This section illustrates the complete cycle for the motivating scenario.
 
\begin{enumerate}[leftmargin=*,topsep=2pt,itemsep=1pt]
\item The user issues the request in Section~\ref{sec:motivation}.
\item The orchestrator calls \texttt{afs list} on each mounted messaging provider (\texttt{/mail}, \texttt{/slack}, \texttt{/teams}, \texttt{/wa}), receiving entries in a uniform schema regardless of the source platform. Each entry carries metadata including source, timestamp, sender, and sensitivity level.
\item The orchestrator merges the entries by timestamp, applies the user's display preferences (concise summaries for Slack and WhatsApp, detailed views for email), and constructs a semantic node tree: a scrollable \texttt{view} containing one child node per message, each with \texttt{src} bound to the original message path and a source-attribution tag in its props. Each message card declares a click event bound to \texttt{exec("/ai/summarize")}.  
\item The orchestrator writes the tree to the AUP session path. The client renders it according to its device capabilities: a card feed on desktop, a compact list on mobile.
\item The user clicks an email card to request a summary. The client emits an event; the UI provider dispatches \texttt{exec("/ai/summarize", \{path: "/mail/inbox/msg-7"\})} to the LLM provider. The provider checks that the entry's sensitivity metadata permits AI processing before proceeding.
\item The LLM provider reads the email content via \texttt{read("/mail/ inbox/msg-7")}, generates a summary, and writes the result to \texttt{"/ui/state/summary"}.
\item The summary node, bound to that path, receives a node-level patch: the summary appears below the message card while the rest of the interface, including any reply the user is composing in another card's input, remains untouched.
\end{enumerate}
 
\noindent Throughout this cycle, all interactions, data access, AI invocation, UI generation, and event handling, occur within a single namespace using uniform operations, eliminating the need for platform-specific APIs or rendering logic. The UI is an integral projection over uniform context.
\section{Discussion}


\textbf{Towards Provider Ecosystem.}
This work assumes that data and capability providers expose their resources through standardized interfaces, introducing additional engineering effort to implement and maintain them.
The central premise is that such standardization enables an ecosystem: once integrated, providers become discoverable, composable, and reusable across agents and applications without repeated integration. This creates network effects, where both new providers and new applications increase the overall value of the system, turning integration into a one-time investment with growing returns.
The approach also supports incremental adoption. Value can be achieved with a small number of providers, and additional ones can be integrated progressively without disrupting existing setups.

\textbf{Toward Trustworthy and Governable Agent Systems.}
In agent-driven systems, real-world data sources often span both personal and professional contexts, making unrestricted access problematic from a trust and governance perspective. This setting requires mechanisms for selective integration of data sources, fine-grained control over permitted operations, and clear visibility into how data is accessed and used. AFS-UI provides a structured approach in which access can be specified at the provider level and resources carry explicit provenance and contextual metadata. These properties support more transparent and traceable system behavior, which is particularly relevant when aggregating information across sources with differing sensitivity levels.

\textbf{From Fixed Applications to Projection Based Semantic Task Spaces.}
This work advances a shift from application-centric interaction toward \textit{semantic task spaces} grounded in a \textit{shared, filesystem-shaped workspace}. Within this formulation, interaction is modeled as a \emph{projection over a persistent context}, where data, capabilities, intermediate states, and traces coexist in a unified, addressable substrate. Interfaces are no longer standalone artifacts, but \textit{situational projections} that organize relevant portions of this shared resource graph for a given task, context, and actor.
This shared workspace provides a common substrate over which multiple forms of interaction can be derived. Graphical interfaces, programmatic APIs, and language- or voice-based interactions can all be understood as alternative projections over the same underlying state. This enables humans and agents to operate within a unified environment, where different input modalities and output forms become interchangeable views rather than separate system layers. As a result, heterogeneous data sources and capabilities can be composed without bespoke integration, while maintaining consistent system state, inspectability, and traceability.

This flexibility introduces trade-offs. Increased inspectability and composability raise cognitive demands, and system stability depends on mechanisms for \textit{semantic identity, provenance, and controlled evolution}. The architecture establishes an operational basis for projection-based interaction environments, demonstrating how semantic task spaces can be instantiated and connected to real data and actions, while highlighting open challenges in usability, stability, and governance.

\section{Related Work}
\label{sec:relatedWork}

\textbf{Agentic Development Frameworks and Platforms.}
A growing ecosystem of frameworks and protocols supports the development of agentic systems for generative UI~\cite{mcp, langchain, claudeCode, mcp-app, mcp-ui, a2ui, ag-ui, json-render}.
At the infrastructure level, the Model Context Protocol (MCP)~\cite{mcp} standardises how tools and data sources are exposed to LLM-based agents. Agent SDKs such as LangChain~\cite{langchain} and systems like Claude Code~\cite{claudeCode} provide abstractions for workflow composition, tool invocation, and execution environments. Skill-based approaches further package reusable capabilities into higher-level units~\cite{vercel-agent-skills, skill, ui-ux-pro-max}. These systems improve usability but remain centered on API-level integration, where tools expose heterogeneous interfaces (C2) and require manual configuration, leaving resource discovery fragmented (C1).

At the UI layer, recent work explores runtime interface generation through different paradigms. Template-based approaches~\cite{mcp-ui, mcp-app} select predefined UI structures, while component-based approaches~\cite{json-render, a2ui} assemble interfaces from reusable components. Other protocols focus on specific aspects of UI delivery: AG-UI~\cite{ag-ui} standardises interaction events, while A2UI and Json-Render define declarative UI specifications. These approaches improve flexibility at the presentation layer but generally assume pre-structured data and tools, resulting in only partial separation between context and projection (C3) and limited support for cross-source composition and substitution (C4).
This work instead targets the underlying interaction model, aiming to unify access, interaction, and composition semantics across resources and interfaces, and thereby address fragmentation (C1), heterogeneity (C2), partial separation (C3), and limited composability (C4).

\textbf{Generative UI Systems for Different Stakeholders.}
Prior work on generative UI spans different stakeholders, including developers and designers~\cite{leviathan2025generative, vercelV0, bolt, lovable, yuan2026towards, chen2025taskartisan, drosos2025dynamic,wang2026ui, park2026bridging, leung2025squire}, and end-users~\cite{hojo2025generativegui, chen2025generative, kim2022stylette, min2025malleable, xu2025duetui}, with each line of work focusing on enhancing different stages of the pipeline.

For developers and designers, generative UI is mainly used for rapid prototyping and implementation. Some approaches generate UI directly from intent in a single step~\cite{leviathan2025generative, vercelV0, bolt, lovable}, while others introduce intermediate representations or multimodal inputs (e.g., screenshots) to better capture design intent~\cite{yuan2026towards, chen2025taskartisan}. Additional work improves controllability by exposing design decisions or providing alternative variants~\cite{wang2026ui, park2026bridging, leung2025squire}. These systems primarily target Stages~2 and~4 and typically assume pre-structured data, components, and design systems, leaving resource resolution (Stage~3) unaddressed.

For end-users, generative UI is studied as an interactive medium for task support and human–AI collaboration. Systems support intent clarification and information exploration~\cite{hojo2025generativegui, chen2025generative, drosos2025dynamic}, personalization and interface reshaping~\cite{kim2022stylette, min2025malleable}, and iterative workflows via user-driven refinement or human–agent collaboration~\cite{cao2025generative, xu2025duetui}. These systems primarily address Stages~1, 2, and~6. However, UI generation (Stages~4--5) often entangles data and specification (C3), and couples data access with interface generation, limiting cross-source recomposition and backend substitution (C4).

Across both lines of work, the task environment is assumed to be predefined and well-structured. Challenges in runtime resource discovery (C1) and heterogeneous operation semantics (C2) remain unresolved, especially in multi-source settings. This work complements prior efforts by supporting Stages~3--5 and part of Stage~6 over heterogeneous, multi-provider environments.

\section{Conclusion}
\label{sec:conclusion}
This paper proposes a \textit{filesystem-based abstraction for agentic generative UI}, reframing UI as a \emph{projection over persistent context}. We identify four challenges and derive five design principles, leading to \textit{AFS-UI}, a unified substrate for organizing heterogeneous data, capabilities, and interaction state for genUI systems.

\bibliographystyle{ACM-Reference-Format}
\bibliography{references}

\end{document}